\documentclass[a4paper,10pt]{article}

\usepackage[T1]{fontenc}
\usepackage{lmodern}
\usepackage[utf8]{inputenc}
\usepackage[margin=1.0in]{geometry}
\usepackage{amsmath,amsfonts,amssymb,mathtools}
\usepackage{hyperref}
\usepackage{graphicx}
\usepackage{xcolor}
\usepackage{subfig}
\usepackage{tikz}
\usepackage{pgfplots}

\numberwithin{equation}{section}

\DeclareMathAlphabet\mathbfcal{OMS}{cmsy}{b}{n}

\newcommand{\beq}{\begin{equation}}
\newcommand{\eeq}{\end{equation}}
\newcommand{\diag}{\text{diag}}
\newcommand{\tr}{\text{Tr}}

\begin{document}

\begin{titlepage}

\begin{flushright}
IFUM-1092-FT
\end{flushright}

\vspace{2.1cm}

\begin{center}
\renewcommand{\thefootnote}{\fnsymbol{footnote}}
{\huge \bf Many accelerating distorted black holes}
\vskip 31mm
{\large {Marco Astorino$^{a}$\footnote{marco.astorino@gmail.com} and Adriano Vigan\`o$^{a,b}$\footnote{adriano.vigano@unimi.it} }}\\

\renewcommand{\thefootnote}{\arabic{footnote}}
\setcounter{footnote}{0}
\vskip 18mm
{\textit{$^{a}$Istituto Nazionale di Fisica Nucleare (INFN), Sezione di Milano \\
Via Celoria 16, I-20133 Milano, Italy}\\
} \vspace{0.3 cm}
{\textit{$^{b}$Universit\`a degli Studi di Milano}} \\
{{\it Via Celoria 16, I-20133 Milano, Italy}\\
}
\end{center}
\vspace{5.6 cm}

\begin{center}
{\bf Abstract}
\end{center}
{
An analytical solution of four-dimensional General Relativity, representing an array of collinear and accelerating black holes, is constructed with the inverse scattering method.
The metric can be completely regularised from any conical singularity, thanks to the presence of an external gravitational field.
Therefore the multi-black hole configuration can be maintained at equilibrium without the need of string or struts.
Some notable subcases such as the accelerating distorted Schwarzschild black hole and the distorted double C-metric are explicitly presented.
The Smarr law and the thermodynamics of these systems is studied.
The Bonnor--Swaminarayan and the Bi\v c\'ak--Hoenselaers--Schmidt particle metrics are recovered, through appropriate limits, from the multi-black holes solutions.
}

\end{titlepage}

\newpage

\tableofcontents

\newpage

\section{Introduction}

Binary and multi-black hole systems are progressively acquiring more and more relevance in the understanding of the large scale structure and interactions of our universe.
That is because gravitational waves detectors can most easily capture gravitational radiation from very massive sources interplay.
These systems generally are not isolated but they belong to larger gravitational structures such as galaxies, or clusters of galaxies which contribute to deform the gravitational field around the black holes and also the behaviour at large distances. 
For these reasons it is worth exploring the possibility to relax the standard asymptotic boundary conditions:
indeed global asymptotic flatness is well known to constitute a big theoretical barrier against the introduction of novel features in the black hole picture, basically because the presence of no-hair theorems in four-dimensional General Relativity.
Global asymptotic flatness also represents a practical obstacle in building an equilibrium, even just meta-stable, multi-black hole configuration, because of the very attractive property of gravity which tends to aggregate and merge matter, when it is not contrasted.

Recently it has been shown in~\cite{Astorino:2021dju,Astorino:2021boj} that the introduction of an external multipolar gravitational background can be useful to analytically build, in pure Einstein gravity, black hole metrics of the Israel--Khan type~\cite{Israel1964}, which can remain at equilibrium without the need of singular matter, such as strings or struts.
In fact these objects plague the spacetime bringing conical defects that violate all reasonable energy conditions or producing $\delta$-like divergences of infinite length.
We would like to avoid these objects because they are theoretically problematic and because there is no clue of their phenomenological existence, so far.

The multipolar gravitational field is not a novel background for (single) black holes.
First introduced by Doroshkevich, Zel'dovich and Novikov~\cite{NovikovZeldovich} in the '50s, these \emph{distorted black holes} have been later studied by several authors including Chandrasekhar~\cite{Chandrasekhar:1985kt} and Geroch--Hartle~\cite{Geroch:1982bv} in the '80s, and several others in more recent times~\cite{breton-manko,Abdolrahimi:2015gea,Fairhurst:2000xh}.
The novelty of our approach is the use of this background to remove conical singularities which usually stem superimposing more black sources.
From a physical point of view it means that the collapse is avoided thanks to the gravitational attraction of the matter surrounding the binary system.
Still, as in the single black hole case, these metrics have to be considered fundamentally as local models.
They are intended to describe the physics of realistic black holes, which are more often surrounded by gravitating matter other than isolated and idealised systems. In fact, nowadays, because of their peculiar astrophysical nature, observed black holes are mainly detected indirectly, thanks to their interaction with bordering matter of the accretion disk.
While these are pure vacuum Einstein manifolds, usually one expects that the full solution may be complemented by some matter distribution which models the presence of a galaxy around the black holes and which provides the desired multipolar expansion.

In this article, inspired by an Ernst's insight in~\cite{ernst}, where he was able to remove the conical singularity of an accelerating black hole thanks to the first non-trivial term (the dipole) of the multipolar expansion of the external gravitational field, we want to push forward his idea and, at the same time, incorporate multi-black hole sources.
More specifically, we want $(i)$ generalise the regularised C-metric constructed by Ernst~\cite{ernst} including the full multipolar expansion of the external gravitational field and $(ii)$ embed an arbitrary number of collinear accelerating black holes into a gravitational background.
The introduction of the acceleration in the deformed black hole scenario, is not just for
sake of generality. But the acceleration parameter often brings with it an extra Killing horizon and a conformal infinity factor, which can improve the local interpretation of these black hole systems embedded in an external gravitational field. 
The idea we pursue in this work was also pioneered by Gibbons, but by regularising with negative mass particles~\cite{Gibbons:1974zd}.

We will focus only on axisymmetric and stationary spacetimes because the solution generating technique we use to build these solution, the inverse scattering method, briefly resumed in section~\ref{sec:ism}, is available only under these assumptions.

The resulting metrics will be analysed in section~\ref{sec:array}, some simple examples will be explicitly worked out in sections~\ref{sec:c-metric} and~\ref{sec:double-c-metric}, and finally some notable limits to well known metrics are unveiled in section~\ref{sec:particles}.

\section{Inverse scattering method}
\label{sec:ism}

The inverse scattering method~\cite{Belinsky:1971nt,Belinsky:1979mh,Belinski:2001ph} allows to superimpose black holes on top of a background spacetime.
For this reason, we start our discussion by presenting the main features of such a solution generation technique and then by setting the so-called ``seed'', that is necessary to construct the accelerating black hole spacetimes.

\subsection{The inverse scattering construction, a brief review}

The inverse scattering method takes advantage of integrability of the Einstein equations for the class of stationary and axisymmetric metrics, which can be described in Weyl coordinates~\cite{Weyl1917} by
\beq
\label{ism-seed}
{ds}^2 = f(\rho,z) ({d\rho}^2 + {dz}^2) + g_{ab}(\rho,z) {dx}^a {dx}^b \, ,
\eeq
where $a,b=0,1$ and $x^0=t$, $x^1=\phi$.
The spacetime~\eqref{ism-seed} has two commuting Killing vectors proportional to $\partial_t$ and $\partial_\phi$.
We postulate that the coordinate $\rho$ is chosen such that $\det g = -\rho^2$.

The vacuum Einstein equations
$R_{\mu\nu} = 0$
can be equivalently written as
\begin{subequations}
\label{einstein}
\begin{align}
\label{eqUV}
U_{,\rho} + V_{,z} & = 0 \, , \\
\label{eqf1}
(\log f)_{,\rho} & = -\frac{1}{\rho} + \frac{1}{4\rho} \tr (U^2-V^2) \, , \\
\label{eqf2}
(\log f)_{,z} & = \frac{1}{2\rho} \tr (UV) \, ,
\end{align}
\end{subequations}
where $U=\rho g_{,\rho} g^{-1}$ and $V=\rho g_{,z} g^{-1}$ are two $2\times2$ matrices.
We note that, by solving equation~\eqref{eqUV} for $g$, one is able to find the function $f$ in quadratures from equations~\eqref{eqf1} and~\eqref{eqf2}.
Thus, the problem of integrating the vacuum Einstein equations is reduced to the problem of constructing the matrix $g$.

One can prove that the integrability condition for the Einstein equations~\eqref{einstein} is equivalent to the linear eigenvalue equations
\beq
\label{schrodinger}
D_1\Psi = \frac{\rho V-\lambda U}{\lambda^2+\rho^2} \Psi \, , \qquad
D_2\Psi = \frac{\rho U+\lambda V}{\lambda^2+\rho^2} \Psi \, ,
\eeq
for the generating matrix $\Psi(\rho,z,\lambda)$,
where the commuting differential operators $D_1$ and $D_2$ are given by
\beq
D_1 \coloneqq \partial_z - \frac{2\lambda^2}{\lambda^2+\rho^2} \partial_\lambda \, , \quad
D_2 \coloneqq \partial_\rho + \frac{2\lambda\rho}{\lambda^2+\rho^2} \partial_\lambda \, ,
\eeq
and $\lambda$ is a complex spectral parameter.

In the inverse scattering technique we pick up a seed solution $(g_0,f_0)$, and then construct a generating matrix $\Psi_0$ which satisfies the linear equations~\eqref{schrodinger}.
Given such a $\Psi_0$, we introduce two functions
\begin{subequations}
\label{solitons}
\begin{align}
\mu_k(\rho,z) & = \sqrt{\rho^2 + (z-w_k)^2} - (z-w_k) \, , \\
\bar{\mu}_k(\rho,z) & = -\sqrt{\rho^2 + (z-w_k)^2} - (z-w_k) \, ,
\end{align}
\end{subequations}
where $w_k$ are arbitrary (complex) constants, which are called poles.
$\mu_k$ and $\bar{\mu}_k$ are called solitons and anti-solitons, respectively, and they satisfy the relation
$\mu_k \bar{\mu}_k = -\rho^2$.

We associate a 2-components vector (the BZ vector) to each (anti-)soliton
\beq
\label{bz}
m_a^{(k)} = m_{0 \, b}^{(k)} \bigl[\psi_0^{-1}(\mu_k,\rho,z)\bigr]_{ba} \, ,
\eeq
where $m_{0 \, a}^{(k)}$ are arbitrary constants.
Defined the matrix
\beq
\Gamma_{kl} = \frac{m_a^{(k)} (g_0)_{ab} m_b^{(l)}}{\rho^2 + \mu_k \mu_l} \, ,
\eeq
a new metric is found by adding $2N$ solitons to $(g_0,f_0)$ as
\begin{subequations}
\label{metric-ph}
\begin{align}
\label{g-ph}
g_{ab} & = \pm \rho^{-2N} \Biggl(\prod_{k=1}^{2N} \mu_k\Biggr) \Biggl[(g_0)_{ab} - \sum_{k,l=1}^{2N} \frac{(\Gamma^{-1})_{kl} L_a^{(k)} L_b^{(l)}}{\mu_k\mu_l}\Biggr] \, , \\
\label{f-ph}
f & = 16 C_f f_0 \rho^{-(2N)^2/2} \Biggl( \prod_{k=1}^{2N} \mu_k^{2N+1} \Biggr) \Biggl[ \prod_{k>l=1}^{2N}(\mu_k-\mu_l)^{-2} \Biggr] \det\Gamma \, ,
\end{align}
\end{subequations}
where
$L_a^{(k)} = m_c^{(k)} (g_0)_{ca}$ and $C_f$ is an arbitrary constant.
The new metric~\eqref{metric-ph} fulfils by construction the Einstein equations~\eqref{einstein}, and also
$\det g=-\rho^2$.

\subsection{Accelerating multipolar gravitational background}

A natural way to obtain the C-metric by means of the inverse scattering technique is to immerse a black hole in an accelerating background.
This means that two solitons have to be added to the Rindler spacetime, which is nothing but Minkowski spacetime adapted to an accelerated observer.
The Rindler spacetime can be expressed in Weyl coordinates in the following way:
\begin{subequations}
\label{acc}
\begin{align}
g_\text{acc} & = \diag\biggl( -\mu_A , \frac{\rho^2}{\mu_A} \biggr) \, , \\
f_\text{acc} & = \frac{\mu_A}{\rho^2 + \mu_A^2} \, ,
\end{align}
\end{subequations}
where $\mu_A = \sqrt{\rho^2 + (z-w_A)^2} - (z-w_A)$ is the soliton which contains the acceleration parameter of the Rindler metric.
A nice parametrisation for the constant $w_A$ is indeed
\beq
w_A = \frac{1}{2A} \,,
\eeq
where $A$ is the acceleration.
The addition of two solitons to~\eqref{acc} gives the standard C-metric, while more (even) solitons allow to construct the accelerating multi-black hole metric discovered by Dowker and Thambyahpillai~\cite{Dowker:2001dg}.

We want to immerse many accelerating black holes in an external gravitational field, hence the natural background is the accelerated version of the external field background presented in~\cite{Astorino:2021boj}.
The external gravitational field is described by the metric
\begin{subequations}
\label{ext}
\begin{align}
g_\text{ext} & = \diag\Biggl[ -\exp\biggr(2\sum_{n=1}^{\infty} b_n r^n P_n \biggr) ,
\rho^2 \exp\biggl(-2\sum_{n=1}^{\infty} b_n r^n P_n \biggr) \Biggr] \, , \\
f_\text{ext} & = \exp\Biggl[
2 \sum_{n,p=1}^\infty \frac{np b_n b_p r^{n+p}}{n+p} \bigl(P_n P_p - P_{n-1} P_{p-1}\bigr)
- 2\sum_{n=1}^{\infty} b_n r^n P_n
\Biggr] \, ,
\end{align}
\end{subequations}
where $r\coloneqq\sqrt{\rho^2+z^2}$ and $P_n=P_n(z/r)$ are the Legendre polynomials.
The parameters $b_n$ are related to the multipole momenta of the external field.
Since the $b_n$ can be chosen at will, the metric~\eqref{ext} can describe a gravitational field generated by a generic distribution of matter~\cite{deCastro:2011zz}.
The non-accelerating multi-black hole solutions immersed in the background~\eqref{ext} were constructed in~\cite{Astorino:2021boj}.

The metric which includes both the acceleration~\eqref{acc} and the external field background~\eqref{ext} is naturally given by
\begin{subequations}
\label{seed}
\begin{align}
g_0 & = \diag\Biggl[ - \mu_A \exp\biggr(2\sum_{n=1}^{\infty} b_n r^n P_n \biggr) ,
\frac{\rho^2}{\mu_A} \exp\biggl(-2\sum_{n=1}^{\infty} b_n r^n P_n \biggr) \Biggr] \, , \\
\begin{split}
f_0 & = \frac{\mu_A}{\rho^2 + \mu_A^2} \exp\Biggl[
2 \sum_{n,p=1}^\infty \frac{np b_n b_p r^{n+p}}{n+p} \bigl(P_n P_p - P_{n-1} P_{p-1}\bigr)
- 2\sum_{n=1}^{\infty} b_n r^n P_n   \\
&\quad + \frac{\rho^2 + \mu_A^2}{\mu_A} \sum_{n=1}^\infty b_n \sum_{l=0}^{n-1} w_A^{n-1-l} \, r^l P_l
\Biggr] \, ,
\end{split}
\end{align}
\end{subequations}
We see that by turning off the external field (i.e.~$b_n=0$), one is left with the Rindler spacetime only.
On the converse, by removing the acceleration, in the limit $w_A\to\infty$, one recovers the background~\eqref{ext}.

Thus, we will take~\eqref{seed} as a background to construct our black hole spacetime.
Since the addition of solitons in the inverse scattering technique is equivalent to the addition of black holes to the seed spacetime~\eqref{seed}, it is quite natural to interpret the resulting metric as a collection of many black holes which are accelerating in an external gravitational field.
Following the discussion in section~\ref{sec:ism}, we need the generating matrix $\Psi_0$ to build a black hole spacetime on top of the background~\eqref{seed}.
The function which satisfies equations~\eqref{schrodinger} generalises the one presented in~\cite{LetelierBrasil}, and it is given by
\beq
\label{psi0}
\Psi_0(\rho,z,\lambda) =
\begin{pmatrix}
- (\lambda-\mu_A) e^{F(\rho,z,\lambda)} & 0 \\
0 & (\lambda+\rho^2/\mu_A) e^{-F(\rho,z,\lambda)}
\end{pmatrix}
\, ,
\eeq
where
\beq
\label{F}
F(\rho,z,\lambda) = 2 \sum_{n=1}^\infty b_n
\Biggl[ \sum_{l=0}^\infty \binom{n}{l} \biggl(\frac{-\rho^2}{2 \lambda}\biggr)^l \biggl(z + \frac{\lambda}{2}\biggr)^{n-l}
- \sum_{l=1}^n \sum_{k=0}^{[(n-l)/2]}
\frac{(-1)^{k+l}2^{-2k-l} n! \lambda^{-l}}{k!(k+l)!(n-2k-l)!}
\rho^{2(k+l)} z^{n-2k-l} \Biggr] \, .
\eeq
Now we can construct the BZ vectors~\eqref{bz}:
we parametrise
$m_0^{(k)}=\bigl(C_0^{(k)},C_1^{(k)}\bigr)$,
where $C_0^{(k)}$, $C_1^{(k)}$ are constants that will be eventually related to the physical parameters of the solution.
The BZ vectors are thus
\beq
m^{(k)} = \biggl( - \frac{C_0^{(k)}}{\mu_k-\mu_0} e^{-F(\rho,z,\mu_k)},
\frac{C_1^{(k)} \mu_0}{\rho^2 + \mu_0\mu_k} e^{F(\rho,z,\mu_k)} \biggr) \, .
\eeq
Depending on the value of $C_0^{(k)}$ and $C_1^{(k)}$, the spacetime will be static or stationary rotating.\\

\section{Array of accelerating black holes in an external gravitational field}
\label{sec:array}

We construct the generalisation of the Dowker--Thambyahpillai solution~\cite{Dowker:2001dg}, which represents an array of collinear accelerating black holes.
The Dowker--Thambyahpillai metric is characterised by the presence of conical singularities, which can not be removed by a fine tuning of the physical parameters without admitting naked singularities.

Given the accelerating background~\eqref{seed} and the generating matrix~\eqref{psi0}, we construct a new solution by adding $2N$ solitons (which correspond to $N$ black holes) with constants
\beq
\label{diagonal}
C_0^{(k)} =
\begin{cases}
1 & k \text{ even} \\
0 & k \text{ odd}
\end{cases}
, \qquad
C_1^{(k)} =
\begin{cases}
0 & k \text{ even} \\
1 & k \text{ odd}
\end{cases}
\, .
\eeq
This choice guarantees a non-rotating metric, which is the one we are interested in.
A different choice for these constants allows the inclusion of the rotation parameter $a$:
for the $k$-th pair of BZ constants one takes
\begin{subequations} \label{rot-Cik}
\begin{align} 
C_0^{(2k-1)} C_0^{(2k)} + C_1^{(2k-1)} C_1^{(2k)} = - \sqrt{m_k^2 - a_k^2} \,, \quad
& C_0^{(2k-1)} C_0^{(2k)} - C_1^{(2k-1)} C_1^{(2k)} = -m_k \frac{1 - A^2 a_k^2}{1 + A^2 a_k^2} \,, \quad \\
C_0^{(2k-1)} C_1^{(2k)} + C_1^{(2k-1)} C_0^{(2k)} = -\frac{2A m_k a_k}{1 + A^2 a_k^2} \,, \quad
& C_0^{(2k-1)} C_1^{(2k)} - C_1^{(2k-1)} C_0^{(2k)} = a_k \,.
\end{align}
\end{subequations}
One can verify that, in the single black hole case and with no external field, i.e. $b_i=0$  for all $i$, the above parametrisation leads to the standard form of the rotating C-metric~\cite{Hong:2004dm}. Likewise the no external gravitational field with the choice (\ref{rot-Cik}), in the multi black hole case, leads to a vacuum multi-Plebanski-Demianski metric\footnote{We do expand explicitly here the full expression of multi-rotating-C-metric, neither with nor without the external gravitational field, because it is quite lengthy, but it can be straightforwardly written down from equations (\ref{metric-ph}) and (\ref{rot-Cik}).}.
Being interested in the phenomenological setting, we have not included the NUT parameter in the above definitions;
however, it is possible to include the NUT charge as well in the inverse scattering formalism~\cite{LetelierSolitons}.

The metric resulting from the diagonal choice~\eqref{diagonal} is
\begin{subequations}
\label{n-acc}
\begin{align}
g_N & = \diag\Biggl[
-\mu_A\frac{\prod_{k=1}^N \mu_{2k-1}}{\prod_{l=1}^N \mu_{2l}} \exp\Biggl(2{\sum_{n=1}^{\infty} b_n r^n P_n}\Biggr),
\frac{\rho^2}{\mu_A} \frac{\prod_{l=1}^N \mu_{2l}}{\prod_{k=1}^N \mu_{2k-1}} \exp\Biggl(-2{\sum_{n=1}^{\infty} b_n r^n P_n}\Biggr)
\Biggr] \, , \\
\begin{split}
f_N & = 16C_f \, f_0 \,
\frac{\mu_A^{2N+1}}{\rho^2+\mu_A^2}
\Biggl( \prod_{k=1}^{2N} \mu_k^{2N+1} \Biggr)
\Biggl( \prod_{k=1}^N \frac{1}{(\mu_A-\mu_{2k})^2} \Biggr)
\Biggl( \prod_{k=1}^N \frac{1}{(\rho^2+\mu_A\mu_{2k-1})^2} \Biggr) \\
&\quad\times \Biggl( \prod_{k=1}^{2N} \frac{1}{\rho^2+\mu_k^2} \Biggr)
\Biggl( \prod_{k=1,l=1,3,\cdots}^{2N-1} \frac{1}{(\mu_k-\mu_{k+l})^2} \Biggr)
\Biggl( \prod_{k=1,l=2,4,\cdots}^{2N-2} \frac{1}{(\rho^2+\mu_k\mu_{k+l})^2} \Biggr) \\
&\quad\times \exp\Biggl[ 2 \sum_{k=1}^{2N} (-1)^{k+1} F(\rho,z,\mu_k) \Biggr] \, .
\end{split}
\end{align}
\end{subequations}
Metric~\eqref{n-acc} is, by construction, a solution of the vacuum Einstein equations~\eqref{einstein},
and it represents a collection of $N$ accelerating black holes, aligned along the $z$-axis, and immersed in the external gravitational field~\eqref{seed}.
Actually, as it usually happens for the C-metrics, the result~\eqref{n-acc} can be interpreted as $N$ \emph{pairs} of black holes which accelerate in two opposite directions~\cite{Griffiths:2006tk}.
However, since the black holes in each pair are causally disconnected, being on two opposite sides of the acceleration horizon and unable to communicate with each other, we restrict our attention to one of the two sides only, and we focus mostly  on the genuine $N$-black hole solution.

We consider real poles $w_k$, since it represents the physically relevant situation.
These constants are chosen with ordering
$w_1<w_2<\cdots<w_{2N-1}<w_{2N}<w_A$
and with parametrisation\footnote{Parametrisation~\eqref{parametrisation} slightly differs from the standard one presented in~\cite{Hong:2003gx}, however it is coherent with \cite{Griffiths:2006tk}. The main advantage of~\eqref{parametrisation} is its close resemblance with the one used for the non-accelerating case~\cite{Astorino:2021boj} and it is more suitable for non accelerating limits.}
\beq
\label{parametrisation}
w_1 = z_1 - m_1\, , \quad w_2 = z_1 + m_1\, , \quad \dotsc \quad
w_{2N-1} = z_N - m_N\, , \quad w_N = z_N + m_N\, , \quad w_A = \frac{1}{2A} \, .
\eeq
The constants $m_k$ represent the black hole mass parameters, $z_k$ are the black hole positions on the $z$-axis and $A$ is the acceleration.

\begin{figure}
\centering
\begin{tikzpicture}
\draw[black,thin] (-10,1) -- (-4,1);
\draw[black,thin] (-10,2) -- (-4,2);
\draw[black,thin] (-2,1) -- (4,1);
\draw[black,thin] (-2,2) -- (4,2);

\draw[black,line width=1mm] (-10,1) -- (-8,1);
\draw[black,line width=1mm] (-8,2) -- (-6,2);
\draw[black,line width=1mm] (-6,1) -- (-4,1);

\draw[black,line width=1mm] (-2,2) -- (0,2);
\draw[black,line width=1mm] (0,1) -- (2,1);
\draw[black,line width=1mm] (2,2) -- (4,2);

\draw [line width=0.7mm,line cap=round,dash pattern=on 0pt off 4\pgflinewidth] (-4,2) -- (-2,2);
\draw [line width=0.7mm,line cap=round,dash pattern=on 0pt off 4\pgflinewidth] (-4,1) -- (-2,1);

\draw[gray,dotted] (-8,2) -- (-8,0);
\draw[gray,dotted] (-6,2) -- (-6,0);
\draw[gray,dotted] (-4,2) -- (-4,0);
\draw[gray,dotted] (-2,2) -- (-2,0);
\draw[gray,dotted] (0,2) -- (0,0);
\draw[gray,dotted] (2,2) -- (2,0);

\draw (-8,-0.2) node{{\small $w_1$}};
\draw (-6,-0.2) node{{\small $w_2$}};
\draw (-4,-0.2) node{{\small $w_3$}};
\draw (-2,-0.2) node{{\small $w_{2N-1}$}};
\draw (0,-0.2) node{{\small $w_{2N}$}};
\draw (2,-0.2) node{{\small $w_A$}};
\draw (4.2,-0.2) node{$z$};

\draw (-10.5,2) node{$t$};
\draw (-10.5,1) node{$\phi$};

\draw[black,->] (-10.5,0) -- (4.2,0);
\end{tikzpicture}
\caption{{\small Rod diagram for the multi-black hole spacetime~\eqref{n-acc}.
The horizons correspond to the timelike rods (thick lines of the $t$ coordinate), while the conical singularities correspond to ``bolts'' where conical singularities can be avoided by imposing an appropriate periodicity on the angular coordinate.}}
\label{fig:rods}
\end{figure}
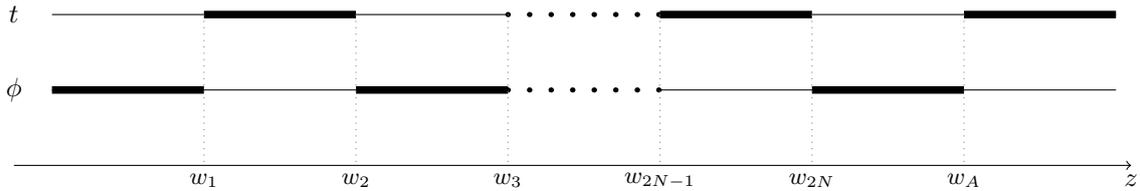

The black hole horizons correspond to the regions
$w_{2k-1}<z<w_{2k}$ ($k=1,\dotsc,N$), while the complementary regions are, in principle, affected by the presence of conical singularities, as happens for the Dowker--Thambyahpillai metric (cf.~Fig.~\ref{fig:rods}).
The metric~\eqref{n-acc} constitutes an extension of the multi-black hole solution presented in~\cite{Astorino:2021boj}, because here we incorporated an additional acceleration horizon which corresponds to the region $z>w_A$ of the spacetime.

\subsection{Behaviour at infinity and acceleration horizon}

The multi-black hole solution~\eqref{n-acc} comes with the black holes curvature singularities which are covered, as usual, by the event horizons.
These singularities do not represent a problem, being the usual ones which are encountered in black hole spacetimes.

However, the external gravitational field metric~\eqref{ext} (and then~\eqref{n-acc}) may be characterised by an unbounded growth of curvature invariants at spatial infinity, which corresponds to $\sqrt{\rho^2+z^2}\to\infty$ in Weyl coordinates.
See~\cite{Abdolrahimi:2015gea} for a detailed study of possible curvature singularity in the domain of outer communication, for the distorted Kerr black hole.
This feature is due to the presence of the sources that generate the external gravitational field, and the latter are thought to be located at large distances from the horizon.
In this respect, as already remarked in~\cite{Astorino:2021dju} and~\cite{Astorino:2021boj}, this kind of metrics have to be considered \emph{local}, in the sense that the global solution would correspond to the matching between the black hole spacetime and an energy-momentum tensor which generates the external field \cite{deCastro:2011zz}.
Hence the behaviour at infinity does not invalidate the physics in proximity of the black holes:
in fact the Smarr law, as well as the first law of thermodynamics, can be achieved for such systems.
Moreover, since these systems are regular in the neighbourhood the horizons, one can even study the second law of thermodynamics in a non-trivial setting~\cite{Astorino:2021dju}.

The novelty of the spacetime~\eqref{n-acc} presented here, is that the curvature unboundedness at infinity is covered by the acceleration horizon given by $z>w_A$.
Thus, the local interpretation of the spacetime is improved in our setting, since the metric is mostly meaningful, between the event and the acceleration horizon.
Being the curvature singularity not directly accessible, the spacetime is completely regular in the physical regions, where observers enjoy the usual metric signature for the manifold and the local model for the distorted multi-black hole system is supposed to hold.\\

\subsection{Regularisation}
\label{sec:regularisation}

The spacetime exhibits conical singularities when the ratio between the length and the radius of small circles around the $z$-axis is different from $2\pi$.
A small circle around the $z$-axis has radius $R=\sqrt{g_{zz}}\rho$ and length $L=2\pi\sqrt{g_{\phi\phi}}$ in Weyl coordinates~\cite{Astorino:2021dju}.
Thus, the regularity condition is nothing but
$L/(2\pi R)\to 1$ as $\rho\to 0$.
It is easy to prove that, for the static and axisymmetric metric~\eqref{ism-seed}, such a condition is equivalent to $\mathcal{P}\equiv f g_{tt}\to 1$ as $\rho\to 0$.
In the case of our multi-black hole metric~\eqref{n-acc}, we can remove the angular defects by choosing the gauge constant $C_f$, and by tuning the external field parameters.

The constant $C_f$ is chosen as
\beq
\begin{split}
\label{n-cf}
C_f & = 2^{4N} \Biggl[ \prod_{i=1}^N (w_{2i}-w_{2i-1})^2 \Biggr]
\Biggl[ \prod_{k=1}^{N-1} \prod_{j=1}^{N-k}
(w_{2k-1} - w_{2k+2j})^2 (w_{2k} - w_{2k+2j-1})^2 \Biggr] \\
&\quad\times \Biggl[ \prod_{l=1}^N (w_A-w_{2l-1})^2 \Biggr]
\exp\Biggl( -2\sum_{n=1}^\infty b_n w_A^n \Biggl)
\, .
\end{split}
\eeq
The quantity $\mathcal{P}=f g_{tt}$ is equal to
\beq
\label{pk}
\mathcal{P}_k =
\Biggl[ \prod_{i=k}^{N-1} \frac{(w_A-w_{2i+1})^2}{(w_A-w_{2i+2})^2} \Biggr]
\Biggl[ \prod_{i=1}^{2k} \prod_{j=2k+1}^{2N}
(w_j-w_i)^{2\,(-1)^{i+j+1}} \Biggr]
\exp\Biggl[ 4\sum_{n=1}^\infty b_n \sum_{j=2k+1}^{2N} (-1)^{j+1} w_j^n \Biggl] \, ,
\eeq
between the $k$-th and $(k+1)$-th black holes (i.e.~$w_{2k}<z<w_{2k+1}$), for $1 \leq k < N$.
In the region $z<w_1$ we find
\beq
\label{p0}
\mathcal{P}_0 =
\Biggl[ \prod_{i=1}^N \frac{(w_A-w_{2i-1})^2}{(w_A-w_{2i})^2} \Biggr]
\exp\Biggl[ 4\sum_{n=1}^\infty b_n \sum_{j=1}^{2N} (-1)^{j+1} w_j^n \Biggl] \, ,
\eeq
while for $w_{2N}<z<w_A$ we simply have
\beq
\mathcal{P}_N = 1 \, ,
\eeq
thanks to our choice of $C_f$.

The expressions~\eqref{pk},~\eqref{p0} provide a system of equations $\mathcal{P}_k=1$, which can be solved for $b_1,\dotsc,b_N$ to completely regularise the spacetime.

\subsection{Smarr law}
\label{sec:smarr}

Let us derive the thermodynamic parameters which appear in the Smarr law.
Firstly, we compute the mass of the spacetime by means of the Komar--Tomimatsu integral~\cite{Komar,Tomimatsu:1984pw}.
The result for the $k$-th black hole (i.e.~the black hole in the interval $w_{2k-1}<z<w_{2k}$) is
\beq
\label{n-mass}
M_k =
\alpha \int_{w_{2k-1}}^{w_{2k}} \rho g_{tt}^{-1} \partial_\rho g_{tt}
= \frac{\alpha}{2} (w_{2k}-w_{2k-1})
= \alpha m_k \, ,
\eeq
where $\alpha$ is a constant which takes into account the proper normalisation of the timelike Killing vector, generator of the horizon, $\xi=\alpha\partial_t$. In general $\alpha$ is not unitary for not asymptotically flat spacetimes.

The black hole entropy is related to the area as $S_k=\mathcal{A}_k/4$, hence
\beq
\label{n-entropy}
S_k =
\frac{1}{4} \lim_{\rho\to0} \int_0^{2\pi} d\phi \int_{w_{2k-1}}^{w_{2k}} dz \sqrt{f g_{\phi\phi}}
= \pi m_k  W
\exp\Biggl[ 2\sum_{n=1}^\infty b_n \sum_{j=2k}^{2N} (-1)^{j+1} w_j^n \Biggl] \, ,
\eeq
where
\beq
\log W = \lim_{\rho\to0} \log\sqrt{f g_{\phi\phi}} =
\log 2 + \sum_{i=1}^{2k-1} \sum_{j=2k}^{2N} (-1)^{i+j+1} \log|w_j-w_i|
+ \sum_{i=2k}^{2N} (-1)^{i+1} \log|w_A-w_i| \, .
\eeq
The temperature is found via the Wick-rotated metric, and the result is
\beq
\label{n-temp}
T_k = \frac{\alpha}{2\pi} \lim_{\rho\to0} \rho^{-1} \sqrt{\frac{g_{tt}}{f}} =
\frac{\alpha}{2\pi} \lim_{\rho\to0} \frac{1}{\sqrt{f g_{\phi\phi}}} =
\frac{\alpha m_k}{2 S_k} \, .
\eeq
It is easy to show, by using~\eqref{n-mass},~\eqref{n-entropy} and~\eqref{n-temp}, that the Smarr law is satisfied:
\beq
\label{n-smarr}
\sum_{k=1}^N M_k = 2 \sum_{k=1}^N T_k S_k \, .
\eeq
The thermodynamics quantities just computed, can be compared with the standard ones in the absence of the external gravitational field~\cite{Gregory:2020mmi}.\\

\section{Accelerating Schwarzschild black hole in an external gravitational field}
\label{sec:c-metric}

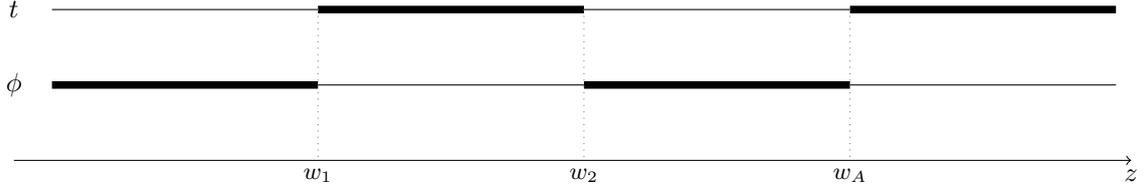
\begin{figure} 
\centering
\begin{tikzpicture}

\draw[black,thin] (-10,2) -- (-6.5,2);
\draw[black,thin] (-6.5,1) -- (-3,1);
\draw[black,thin] (-3,2) -- (0.5,2);
\draw[black,thin] (0.5,1) -- (4,1);

\draw[black,line width=1mm] (-10,1) -- (-6.5,1);
\draw[black,line width=1mm] (-6.5,2) -- (-3,2);
\draw[black,line width=1mm] (-3,1) -- (0.5,1);
\draw[black,line width=1mm] (0.5,2) -- (4,2);

\draw[gray,dotted] (-6.5,2) -- (-6.5,0);
\draw[gray,dotted] (-3,2) -- (-3,0);
\draw[gray,dotted] (0.5,2) -- (0.5,0);

\draw (-6.5,-0.2) node{{\small $w_1$}};
\draw (-3,-0.2) node{{\small $w_2$}};
\draw (0.5,-0.2) node{{\small $w_A$}};
\draw (4.2,-0.2) node{$z$};

\draw (-10.5,2) node{$t$};
\draw (-10.5,1) node{$\phi$};

\draw[black,->] (-10.5,0) -- (4.2,0);

\end{tikzpicture}
\caption{{\small Rod diagram for the C-metric immersed in the external field~\eqref{1-acc}.
We notice that the presence of the external field does not affect the rod diagram, i.e.~the structure of the poles.
Being the rods defined by the poles $w_n$ (see~\cite{Harmark:2004rm}), we obtain the usual C-metric diagram.}}
\label{fig:c-metric}
\end{figure}

The first specialization of the multi-source metric proposed above it is worth discussing is the single black hole case, i.e.~$N=1$.
The very first prototype of these kind of metrics were built by Ernst~\cite{ernst} with the aim of regularising the conical singularity of the C-metric, through the presence of an external gravitational field possessing only the first term of the external multipolar expansion, the dipole\footnote{The zeroth-order of the external multipolar expansion is just a constant that can be reabsorbed.
While the dipole term in the standard internal multipolar expansion is often, under certain assumptions, washed away thanks to a coordinate shift to the center of mass, in this external multipolar expansion this change of reference cannot erase the dipole contribution.}.
The $N=1$ characterization of the metric~\eqref{n-acc} represents the full multipolar expansion of the external gravitational field with respect to the Ernst solution.
Moreover our metric carries an extra parameter $z_1$ which describes the position of the black hole with respect to the multipoles\footnote{This differs with respect to the Ernst metric, which considered a fixed position for the black hole in the center of the coordinate system, that is for $-m/A<z<m/A$.}.
The metric is quite simple and can be written as~\eqref{ism-seed} with
\begin{subequations}
\label{1-acc}
\begin{align}
g_\text{1} & = \diag\Biggl[ - \frac{\mu_1  \mu_A}{\mu_2}  \exp\biggr(2\sum_{n=1}^{2} b_n r^n P_n \biggr) ,
\rho^2 \frac{\mu_2}{\mu_1 \mu_A} \exp\biggl(-2\sum_{n=1}^{2} b_n r^n P_n \biggr) \Biggr] \, , \\
f_\text{1} & = -\frac{ 16 C_f f_0 w_1^2 \mu_1^3 \mu_2^3 \exp \left[ 2F(\rho,z,\mu_1) - 2F(\rho,z,\mu_2) \right] }{(\rho^2 + \mu_1^2)(\rho^2 + \mu_2^2)(\mu_A-\mu_1)^2(\mu_A-\mu_2)^2(\mu_1-\mu_2)^2} \, ,
\end{align}
\end{subequations}
where the sum in $F$~\eqref{F} is limited to the second order. The rod diagram remains the same of the standard C-metric, since the poles are not affected by the presence of the external gravitational field, as can be appreciated in figure \ref{fig:c-metric}.

The limit to the standard C-metric in spherical coordinates $(t,r,\theta,\varphi)$
\beq
\label{c-metric}
\begin{split}
{ds}^2 & = \frac{1}{(1 + A r \cos \theta)^2} \biggl[ -\biggl(1-\frac{2m}{r} \biggr) \bigl(1-A^2r^2\bigr) {dt}^2 + \frac{dr^2}{\bigl(1-\frac{2m}{r} \bigr)\bigl(1-A^2r^2 \bigr)} \\
&\quad + \frac{r^2 {d\theta^2}}{(1+2Am\cos\theta)} + r^2 (1+2Am\cos\theta) \sin^2 \theta {d\phi}^2 \biggr] \,,
\end{split}
\eeq
is obtained thanks to the rescaling of time $t\to \sqrt{A}\, t$, azimuthal angle $\phi \to \phi/\sqrt{A}$ and the following change of coordinates
\beq
\rho = \frac{\sin\theta \sqrt{r(r-2m)(1-A^2r^2)(1+2Am\cos\theta)}}{(1+Ar\cos\theta)^2} \,, \quad
z = z_1 + \frac{(Ar+\cos\theta)(r-m+Amr\cos\theta)}{(1+Ar\cos\theta)^2} \,,
\eeq
vanishing $z_1$ and the whole multipolar expansion:
$b_n=0, \,\, \forall n$.
To obtain exactly Eq.~\eqref{c-metric}, one has to fix the solitonic constants in terms of the physical quantities
\beq
\label{acc-wi}
w_1 =  z_1 - m \, , \quad
w_2 =  z_1 + m \, , \quad
w_A = \frac{1}{2A} \, , \quad
C_f = \frac{1}{8A^3} \, .
\eeq
These coordinates and parametrization are also useful for describing the accelerating and distorted black hole metric.

However, in order to remove the two conical singularities, which are generally present in the $z$-axis of the single accelerating black hole above, some constants have to be properly constrained, as explained in section~\ref{sec:regularisation}. 
For simplicity, in this section we will focus on the first two terms of the multipolar expansion:
$b_n=0, \, \forall n > 2$.
In this case we can explicitly write down the values of the physical parameters which regularise the metric for $z \in (z_1+m,1/2A)$ and $z\in (-\infty,z_1-m)$ respectively:
\begin{align}
\label{reg-acc-Cf}
\begin{split}
C_f & = -\frac{(w_1-w_2)^2(w_1-w_A)^2}{8w_1^2} \exp\bigl[ -2w_A(b_1+b_2w_A)  \bigr] \\
& = -\frac{m + 2 A m (m-z_1)^2}{8A^2 (m-z_1)^2} \exp \biggl(-\frac{2 A b_1+b_2}{2A^2} \biggr) \,,
\end{split}
\\
\label{reg-acc-b1}
b_1 &= \frac{2 b_2 (w_2^2-w_1^2) + \log(\frac{w_2-w_A}{w_1-w_A})}{2(w_1-w_2)} = -2 b_2 z_1 - \frac{1}{4m} \log \biggl[\frac{1-2A(m+z_1)}{1+2A(m-z_1)} \biggr] \,.
\end{align}
Note that the $w_i$ remain defined as in~\eqref{acc-wi}, while in the presence of the external gravitational field $C_f$ can be upgraded to eliminate the conicity,  as done in~\eqref{reg-acc-Cf}.
We remark that this external multipolar distortion is usually designed to model black holes locally.
Nevertheless, thanks to the above regularisation, the spacetime remains completely regular in the physical regions.
In fact, between the event horizon and the accelerating horizon, where the signature is $(-+++)$, the metric is free from both conical and curvature singularities, for any finite value of the accelerating parameter $A$ satisfying the rod ordering relation $z_1-m<\frac{1}{2A}$.

To have a better intuition on the physics introduced by the external external field, it is instructive to analyse the weak field limit of the accelerating metric~\eqref{1-acc}, that is when the black hole mass parameter $m$ is small.
In this case we can appreciate the contribution of the external multipoles on the acceleration imparted to inertial observer, which we consider located in the origin of the (spherical) coordinates for simplicity. 
The tetra-dimensional timelike worldline of an observer with proper time $\lambda$ and constant radial $\bar{r}$ and polar coordinates $\bar{x}=\cos\bar{\theta}$ is 
\beq
y^\mu(\lambda) =
\Biggl(
\frac{1+A\bar{r}\bar{x}}{\sqrt{1-A^2\bar{r}^2}}
e^{-\frac{\bar{r}}{2(1+A\bar{r}\bar{x})} \bigl[ 2b_2\bar{r} (A \bar{r}+\bar{x})^2 +2 b_1 (A \bar{r}+\bar{x}) (1+ A \bar{r}\bar{x})^2-b_2 \bar{r} (1-A^2\bar{r}^2)(1-\bar{x}^2) \bigr]} \lambda,
\bar{r},
0,
0,
\Biggr)
\,.
\eeq
This choice fulfils the normalisation property of the four-velocity, $u_\mu u^\mu=-1$, where $u^\mu \coloneqq dy^\mu(\lambda)/d\lambda$.
The absolute value of the four-acceleration, $a^\mu \coloneqq (\nabla_\nu u^\mu)u^\nu$, for this observer is given by
\beq
\label{acc-bi}
|a| = \sqrt{a_\mu a^\mu}\  \Big|_{\bar{r}=0} =
| A -b_1 | \exp\biggl(-\frac{b_2+2Ab_1}{4A^4} \biggr)  \, .
\eeq 
Note that, because $a^\mu u_\mu=0$, $|a|$ corresponds also to the magnitude of the three-acceleration in the rest frame of the observer,
the external gravitational field has a non-trivial role in the background acceleration quantified by~\eqref{acc-bi}.
In the vanishing multipoles limit, $b_i=0$, the standard C-metric acceleration $|a|=A$ is retrieved from~\eqref{acc-bi}.

The non-relativistic limit, i.e.~for small values of the accelerating parameter $ A \approx 0 $, of the regularising condition~\eqref{reg-acc-b1} can provide some further understanding of the multipolar deformations.
In fact in this approximation we expect to retrieve a Newtonian picture:
the force felt by a massive monopole in an uniform gravitational field is
\beq
\label{approx}
m A \approx  \frac{1- e^{ -4 m (b_1+2b_2 z_1)}}{4} \,.
\eeq
One obtains a very simple expression when the external field is weak, i.e.~$b_1\approx b_2\approx 0$:
in such a case the exponential in~\eqref{approx} can be expanded and
\beq
m A \approx m (b_1 + 2 b_2 z_1) .
\eeq
The last equation is nothing but the Newton law $\vec{F}=m\vec{a}$, hence $b_1 + 2 b_2 z_1$ is interpreted, in the Newtonian limit, as a constant external gravitational field strength.
We see that the regularisation condition has a nice and physically transparent limit, which is consistent with the analysis performed by Bonnor~\cite{bonnor1988exact} on the Ernst solution.

The metric described in this section has interesting applications in the realm of black hole pair creation.
Usually it is speculated that in the presence of a strong electromagnetic field the C-metric can describe a couple of casually disconnected and charged black holes (eventually rotating ~\cite{Astorino:2013xxa}) popping-out from vacuum and accelerating away~\cite{Gibbons:1986cq,strominger,hawking}.
In that picture the regularising interaction between the electric charge of the black hole and the background electromagnetic Bonnor--Melvin universe~\cite{bonnor,melvin} reveals to be crucial.
On the other hand, the metric presented in this section provides the regularisation by means of the external gravitational field.
Therefore, in this setting the pair creation of black holes is fostered by the energy of the external gravitational field.
Thus this picture seems to be more phenomenological, because it does not require the black holes to be constitute by charged matter, an occurrence that appears outside empirical observations, at the moment.
In fact the metric~\eqref{1-acc} allows the pair creation of neutral black holes in a gravitational background.
The rate of the pair creation is proportional to the intensity of the external gravitational field, but
its computation is outside the scope of this paper and will be done elsewhere~\cite{pair-bh-grav-field}. 

The thermodynamics quantities and the Smarr law follow directly from the general multiple case presented in the section~\ref{sec:smarr}.
Otherwise, the first law of black hole thermodynamics can be retrieved as a trivial specialisation of the double configuration studied in the next section, when one of the two masses of the double configuration vanishes.

\section{Double C-metric in an external gravitational field}
\label{sec:double-c-metric}

The accelerating external gravitational background provides us a second interesting opportunity to generalise the solution describing a binary black hole system at equilibrium, as presented in~\cite{Astorino:2021dju,Astorino:2021boj}.
This extension not only represents an enrichment of the physical model features, but it provides also a mechanism to protect the most physical region of the black hole against the unbounded growth of the scalar curvature invariants at spatial infinity.
In fact, for finite values of the radial coordinate $r$, an observer will encounter the accelerating Killing horizon and the conformal infinity before reaching spatial infinity, further enforcing the local nature of the model.

As explained in section~\ref{sec:array}, the metric can be analytically generated for the whole multipolar series, which includes infinite independent terms, each one with its independent integration constant.
However in this section we work out explicitly a simple example on a truncated multipolar expansion, namely keeping only the dipole and quadrupole deformations.
Indeed these two quantities are sufficient to regularise the metric without constraining the proper physical parameters of the black holes configuration\footnote{$F$ in $f_2$ is defined according to~\eqref{F}, but only up to the second order.}. That's because the number of constraints on the physical parameters of the metric coincides with the number of casually connected black holes, and we are now considering a double accelerating c-metric. In this case the two blocks of the spacetime metric can be written as follows\footnote{A Mathematica worksheet with this metric can be found among the files of this arXiv paper and on the web-page \href{ https://sites.google.com/site/marcoastorino/}{https://sites.google.com/site/marcoastorino/}.}
\begin{subequations}
\label{2-acc}
\begin{align}
g_2 & = \diag\Biggl[ - \frac{\mu_1 \mu_3 \mu_A}{\mu_2 \mu_4}  \exp\biggr(2\sum_{n=1}^{2} b_n r^n P_n \biggr) ,
\frac{\rho^2 \mu_2 \mu_4}{\mu_1 \mu_3 \mu_A} \exp\biggl(-2\sum_{n=1}^{2} b_n r^n P_n \biggr) \Biggr] \, , \\
f_2 & = \frac{ f_1 w_3^2 \mu_1^2 \mu_2^2 \mu_3^5 \mu_4^5 \exp [2F(\rho,z,\mu_3) - 2F(\rho,z,\mu_4)] }{(\rho^2 + \mu_1\mu_3)(\rho^2 + \mu_3^2)(\rho^2+\mu_2\mu_4)(\rho^2 + \mu_4^2)(\mu_2 - \mu_3)^2(\mu_A-\mu_3)^2(\mu_1-\mu_4)^2(\mu_3-\mu_4)^2(\mu_A-\mu_4)^2} \, ,
\end{align}
\end{subequations}
where $f_1$ is the single-black hole value for the non-Killing elements of the metric encountered in previous section.
We have to properly tune three parameters\footnote{Two of these parameters, the $b_i$, are relate to physical quantities, while $C_f$ is a gauge constant of the metric.} of the solution to obtain a metric devoid of angular defects, one for each sector, on the $z$-axis, between the timelike rods of Fig.~\ref{fig:rods}.
A possible choice in terms of the poles $w_i$ is\footnote{A more physical parametrisation is achieved with the constants of~\eqref{parametrisation}.}
\begin{align}
\label{Cf-2}
C_f &= \frac{8 e^{-2w_A(b_1+b2w_A)}}{w_1^2 w_3^2}(w_1-w_2)^2(w_2-w_3)^2(w_1-w_4)^2(w_3-w_4)^2(w_1-w_A)^2(w_3-w_A)^2 \,, \\
\label{b1-2}
b_1  &= \frac{4b_2(w_2^2-w_1^2+w_4^2-w_3)+\log\Bigl[\frac{(w_A-w_2)^2(w_A-w_4)^2}{(w_A-w_1)^2(w_A-w_3)^2} \Bigr]}{4(w_1-w_2+w_3-w_4)} \,, \\
\label{b2-2}
b_2  &= \frac{2(w_1-w_2+w_3-w_4)\log\Bigl[\frac{(w_1-w_3)(w_2-w_4)(w_4-w_A)}{(w_2-w_3)(w_1-w_4)(w_3-w_A)} \Bigr] +(w_4-w_3) \log \Bigl[\frac{(w_1-w_3)(w_2-w_4)(w_4-w_A)}{(w_2-w_3)(w_1-w_4)(w_3-w_A)}  \Bigr]}{4(w_1-w_2)(w_3-w_4)(-w_1-w_2+w_3+w_4)} \,.
\end{align}
Physically this choice can be interpreted as the specific deformation of the external gravitational field multipoles required to support a generic binary configuration.
Other possible choices may have different physical interpretation:
for instance, fixing the position of the holes through $z_i$ instead of $b_1$ to eliminate the conical singularities, corresponds in adjusting the position of the black holes in a given multipolar configuration. 
Henceforth we will considered the value of the three parameters $b_i$ and $C_f$ constrained as in~\eqref{Cf-2}-\eqref{b1-2} to ensure the spacetime to be free from angular defects anywhere.

The equilibrium is achieved for finite proper distance of the two black hole sources, as it can be checked from
\beq
\ell = \int_{w_2}^{w_3} dz \sqrt{g_{zz}(\rho,z)} \Big|_{\rho=0} < \infty \, .
\eeq

\begin{figure}[h]
\centering
\hspace{-0.5cm}
\includegraphics[scale=0.36]{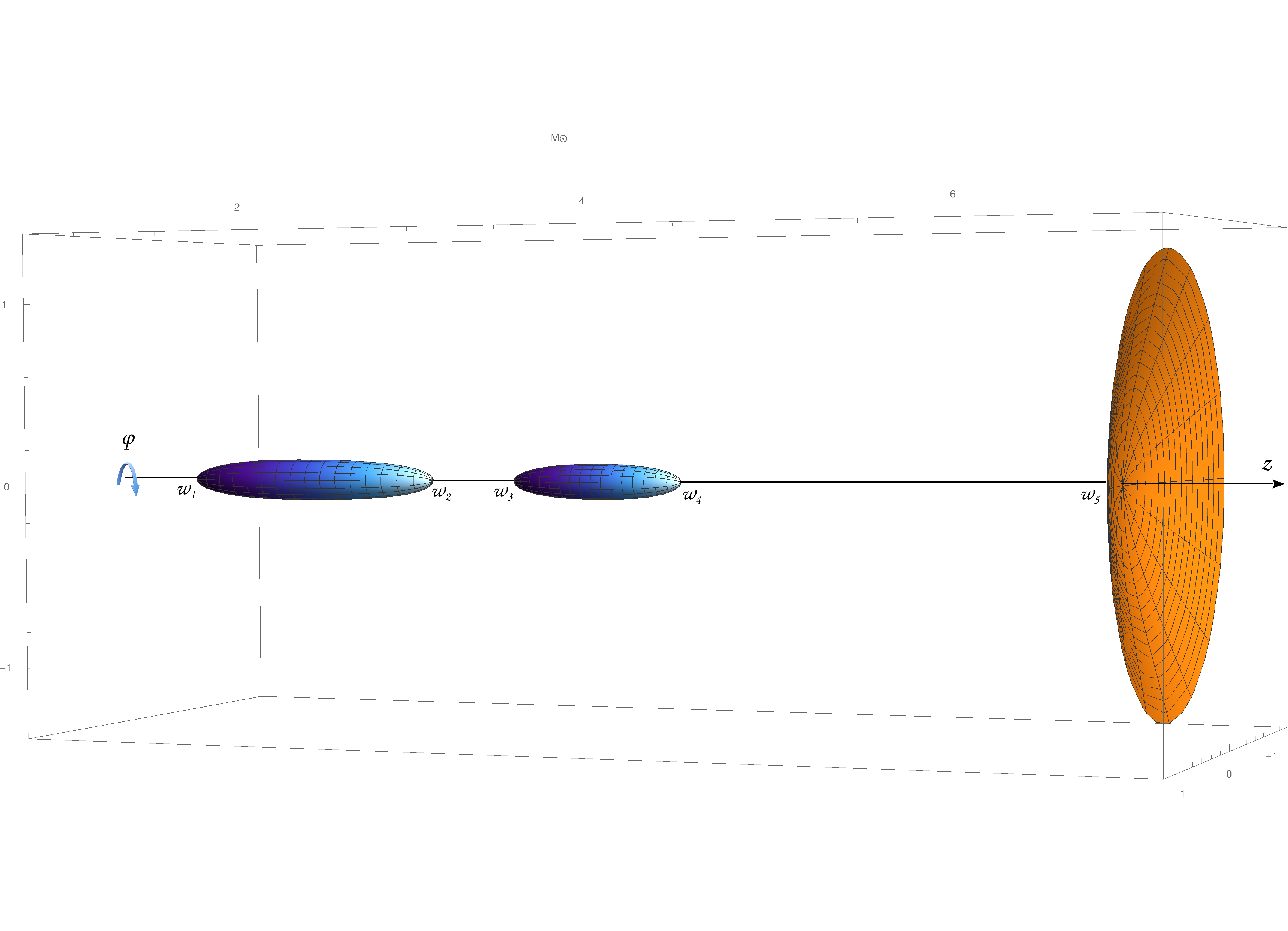}
\caption{\small Embedding diagram in $\mathbb{E}^3$ of the binary accelerating black hole horizons for the parametric values $w_1=1$, $w_2=2.5$, $w_3=3$, $w_4=4$, $w_A=6.5$ expressed in Solar mass units $M_\odot$.
The blue surfaces are the event horizons of the binary black system, while the yellow surface represents a small cylindrical section of the the accelerating horizon.
All the horizon are smooth surfaces thanks to constraints \eqref{Cf-2},~\eqref{b1-2},~\eqref{b2-2}.
The accelerating horizon for $z<0$ is not visible because in that sector it is shifted towards conformal infinity, as usually occurs for C-metrics.}
\label{fig:embedding3}
\end{figure}

In figure \ref{fig:embedding3} the regularised black holes event and accelerating horizons can be appreciated. It is also apparent how the Rindler horizon is deformed, on the axis of symmetry, by the tidal forces of the binary system.

\subsection{Thermodynamics}

The mass of each member of the double configuration can be evaluated integrating on their respective rod, as done for the general case~\eqref{n-mass}:
$M_i = \alpha m_i $, thus the total mass is given by $M = M_1 + M_2$.
The total entropy of the system is taken as the quarter of the two black hole surfaces, as described in section~\ref{sec:smarr}:
$S = S_1 + S_2$,
where
\begin{subequations}
\begin{align}
\begin{split}
S_1 & = \frac{\pi (w_1-w_2)^2(w_1-w_4)(w_3-w_A)}{2 (w_1-w_3)(w_2-w_A)(w_5-w_A)} e^{-2b_1(w_2-w_3+w_4)-2b_2(w_2^2-w_3^2+w_4^2)} \\
& = \frac{4 A e^{-2b_1(m_1+2m_2+z_1)-2b_2(m_1^2+2m_1z_1+z_1^2+4m_2z_2)}m_1^2\pi[1+2A(m_2-z_2)](m_1+m_2-z_1+z_2)}{[-1+2A(m_1+z_1)](m_1-m_2-z_1+z_2)(-1+2A(m_2+z_2))}  \,,
\end{split}
\\
\begin{split}
S_2 & = \frac{\pi (w_3-w_4)^2(w_4-w_1)}{2(w_2-w_4)(w_4-w_A)}  e^{-2w_4(b_1+b_2w_4)} \\ 
& = - \frac{4Ae^{-2(m_2+z_2)(b_1+b_2(m_2+z_2))}m_2^2\pi(m_1+m_2-z_1+z_2)}{(-m_1+m_2-z_1+z_2)[-1+2A(m_2+z_2)]} \,.
\end{split}
\end{align}
\end{subequations}
The two horizon temperatures, computed as in~\eqref{n-temp}, simply result $T_i =M_i/(2S_i)$, fulfilling straightforwardly the Smarr law both for each single black source and for the binary configuration.

The first law of black hole thermodynamics can also be verified, once the normalisation of the timelike Killing vector is chosen to make the mass of the system integrable and continuously connected with the known cases, i.e.~the non-accelerating configuration and the single black hole configuration.
The Christodoulou--Ruffini mass formula~\cite{Christodoulou:1972kt} suggests the use of an integrating factor $\alpha$ such that 
\beq
\sum_{i=1}^2 M_i = \sum_{i=1}^2 \sqrt{\frac{S_i}{4\pi}} \,.
\eeq
It can be verified that this occurs whenever 
\beq
w_A = \frac{e^{2b_1(w_2-w_3+w_4)+2b_2(w_2^2-w_3^2+w_4^2)}w_2(w_3-w_2)+e^{2w_4(b_1+b_2w_4)}w_3(w_2-w_4)}{e^{2b_1(w_2-w_3+w_4)+2b_2(w_2^2-w_3^2+w_4^2)}(w_3-w_2)+e^{2w_4(b_1+b_2w_4)}(w_2-w_4)} \, .
\eeq
The resulting value of the integrating factor is
\beq
\alpha = e^{-w_4(b_1+b_2w_4)} \sqrt{\frac{w_4-w_2}{2(w_4-w_2)(w_5-w_4)}} \, . 
\eeq
Then the first law of black hole thermodynamics holds for each member of the black hole configuration as follows
\beq
\delta M_i =  T_i \, \delta S_i \, .
\eeq
In the presence of thermodynamic equilibrium between the two horizons, $T_1=T_2$, which can be achieved constraining another integrating constant of the solution (e.g.~$w_4$), a first law for the whole black hole configuration can be written
\beq
\delta M = T \, \delta S \, .
\eeq

\section{Accelerating particles in an external gravitational field}
\label{sec:particles}

There exist many particle-like solutions in General Relativity that have been extensively studied over the years.
These solutions are related to the Curzon--Chazy family of metrics~\cite{Curzon,Chazy}, and they represent the gravitational field generated by point-like particles.
The particles themselves are nothing but naked singularities\footnote{The structure of these curvature singularities is quite complicated and depends on the direction one approaches it. See~\cite{Griffiths:2006tk} and references therein.}.
It is quite easy to construct a metric that contains a collection of many Curzon--Chazy particles:
these multi-particle solutions are affected by the presence of conical singularities~\cite{Einstein:1936fp}, as expected on physical grounds.

The accelerated version of two Curzon--Chazy particles was found by Bonnor and Swaminarayan~\cite{bonnor1964exact}:
in such a solution the particles are accelerated by two cosmic strings\footnote{Actually, the conical singularities disappear when one of the two particles has negative mass.}.
The metric, in the case of a single accelerating particle, was later regularised, following the lines of~\cite{ernst}, by Bi\v c\'ak, Hoenselaers and Schmidt~\cite{Bicak} with the introduction of an external field, by which they were able to remove the conical singularities.
Moreover, they showed that their external field, that actually corresponds to our multipolar expansion~\eqref{ext} when $b_n=0$ for $n\geq2$, can be obtained by sending to infinity the second particle of the Bonnor--Swaminarayan solution.
This corroborates the idea that the field background is generated by sources located at infinity.

It is worth mentioning an alternative approach, pursued by Gibbons~\cite{Gibbons:1974zd}, who managed to regularise an accelerating black hole by means of a negative-mass Curzon--Chazy particle.

We will show that the generalisation of the Bi\v c\'ak--Hoenselaers--Schmidt solution to $N$ particles and with the generic gravitational field~\eqref{acc} can be achieved by means of an appropriate limit of our multi-black hole solution~\eqref{n-acc}.

\subsection{The limit to the Bonnor--Swaminarayan metric}

Let us begin by considering the limit to the Bonnor--Swaminarayan solution, i.e.~two accelerating particles with no external gravitational field.
We consider then the metric~\eqref{n-acc} with $b_n=0$ for all $n$.
To clarify the limit, we specialise to the case $N=2$, nonetheless the generalisation any $N$ is straightforward.

It is useful to rewrite the metric~\eqref{n-acc} in the canonical Weyl form
\beq
\label{weyl}
{ds}^2 = -e^{2\psi(\rho,z)} {dt}^2 + e^{-2\psi(\rho,z)} \bigl[ e^{2\gamma(\rho,z)} \bigl( {d\rho}^2 + {dz}^2 \bigr) + \rho^2 {d\phi}^2 \bigr] \,,
\eeq
and to work with the potential
\beq
\psi = \frac{1}{2} \log\biggl(\frac{\mu_1}{\mu_2}\biggr) + \frac{1}{2} \log\biggl(\frac{\mu_3}{\mu_4}\biggr) + \frac{1}{2} \log\mu_A \,.
\eeq
Noticing that
\beq
\frac{\mu_k}{\mu_{k+1}} = \frac{\mu_k - \bar{\mu}_{k+1}}{\mu_{k+1} - \bar{\mu}_k} =
\frac{R_k + R_{k+1} - 2m_k}{R_k + R_{k+1} + 2m_k} \,,
\eeq
where
\beq
R_k = \sqrt{\rho^2 + (z + m_k - z_k)^2} \,, \quad
R_{k+1} = \sqrt{\rho^2 + (z - m_k - z_k)^2} \,,
\eeq
we can write
\beq
\psi = \frac{1}{2} \log\biggl(\frac{R_1 + R_2 - 2m_1}{R_1 + R_2 + 2m_1}\biggr) + \frac{1}{2} \log\biggl(\frac{R_3 + R_4 - 2m_2}{R_3 + R_4 + 2m_2}\biggr) + \frac{1}{2} \log\mu_A \,.
\eeq
One recognises two Schwarzschild potentials (the first two terms) and the Rindler potential (the last term).
Indeed, at the level of the Weyl potential a superposition principle holds;
the non-linearity is encoded in the function $\gamma$, that we do not explicitly write here.

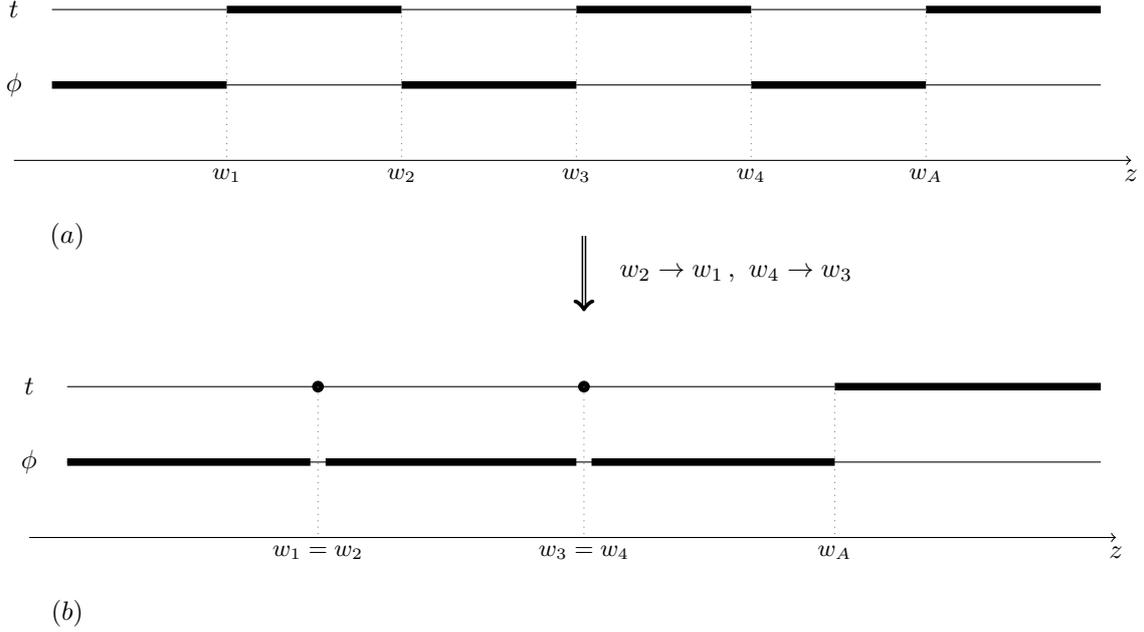
\begin{figure}
\centering
\begin{tikzpicture}

\draw[black,thin] (-10,2) -- (-7.7,2);
\draw[black,thin] (-7.7,1) -- (-5.4,1);
\draw[black,thin] (-5.4,2) -- (-3.1,2);
\draw[black,thin] (-3.1,1) -- (-0.8,1);
\draw[black,thin] (-0.8,2) -- (1.5,2);
\draw[black,thin] (1.5,1) -- (3.8,1);

\draw[black,line width=1mm] (-10,1) -- (-7.7,1);
\draw[black,line width=1mm] (-7.7,2) -- (-5.4,2);
\draw[black,line width=1mm] (-5.4,1) -- (-3.1,1);
\draw[black,line width=1mm] (-3.1,2) -- (-0.8,2);
\draw[black,line width=1mm] (-0.8,1) -- (1.5,1);
\draw[black,line width=1mm] (1.5,2) -- (3.8,2);

\draw[gray,dotted] (-7.7,2) -- (-7.7,0);
\draw[gray,dotted] (-5.4,2) -- (-5.4,0);
\draw[gray,dotted] (-3.1,2) -- (-3.1,0);
\draw[gray,dotted] (-0.8,2) -- (-0.8,0);
\draw[gray,dotted] (1.5,2) -- (1.5,0);

\draw (-7.7,-0.2) node{{\small $w_1$}};
\draw (-5.4,-0.2) node{{\small $w_2$}};
\draw (-3.1,-0.2) node{{\small $w_3$}};
\draw (-0.8,-0.2) node{{\small $w_4$}};
\draw (1.5,-0.2) node{{\small $w_A$}};
\draw (4.2,-0.2) node{$z$};

\draw (-10.5,2) node{$t$};
\draw (-10.5,1) node{$\phi$};

\draw[black,->] (-10.5,0) -- (4.2,0);

\draw (-9.8,-1) node{$(a)$};

\draw[black,line width=0.2mm,double,->] (-3,-1) -- (-3,-2);
\draw (-1,-1.5) node{$w_2\to w_1 \,, \,\, w_4\to w_3$};

\draw[black,thin] (-9.8,-3) -- (3.8,-3);
\draw[black,thin] (-9.8,-4) -- (3.8,-4);

\draw[black,line width=1mm] (-9.8,-4) -- (-6.6,-4);
\draw[black,line width=1mm] (-6.4,-4) -- (-3.1,-4);
\draw[black,line width=1mm] (-2.9,-4) -- (0.3,-4);
\draw[black,line width=1mm] (0.3,-3) -- (3.8,-3);

\filldraw [black] (-6.5,-3) circle (2pt);
\filldraw [black] (-3,-3) circle (2pt);

\draw[gray,dotted] (-6.5,-3) -- (-6.5,-5);
\draw[gray,dotted] (-3,-3) -- (-3,-5);
\draw[gray,dotted] (0.3,-3) -- (0.3,-5);

\draw (-6.5,-5.2) node{{\small $w_1=w_2$}};
\draw (-3,-5.2) node{{\small $w_3=w_4$}};
\draw (0.3,-5.2) node{{\small $w_A$}};
\draw (4,-5.2) node{$z$};

\draw (-10.3,-3) node{$t$};
\draw (-10.3,-4) node{$\phi$};

\draw[black,->] (-10.3,-5) -- (4,-5);

\draw (-9.8,-6) node{$(b)$};

\end{tikzpicture}
\caption{{\small Rod diagrams for $(a)$ the double C-metric in an external field~\eqref{2-acc} and for $(b)$ the Bonnor--Swaminarayan metric~\eqref{bonnor}.
We see that the limit considered in the main text $w_{2k}\to w_{2k-1}$ corresponds to shrinking the timelike rods representing the event horizons.
The horizons disappear through that limit, and the resulting objects are naked singularities which are represented by points in the rod diagram.}}
\label{fig:particles}
\end{figure}

Now we consider the limit in which the finite timelike rods of Fig.~\ref{fig:rods} are pinched to a point, i.e.~when $w_{2k}\to w_{2k-1}$:
this is equivalent to consider $m_k\to 0$.
We expand for small $m_k$ to order $O\bigl(m_k^2\bigr)$, to find
\beq
\psi \approx -\frac{m_1}{\sqrt{\rho^2 + (z - z_1)^2}} - \frac{m_2}{\sqrt{\rho^2 + (z - z_2)^2}} + \frac{1}{2} \log\mu_A \,.
\eeq
Again, we recognise the various terms in the last expression:
the first two are Curzon--Chazy potentials and represent point-like particles, while the last term is still the Rindler one.
Then it is natural to interpret the potential as the one corresponding to two accelerating particles.

We cast the potential in the usual Bonnor--Swaminarayan by performing the change of coordinate
$\bar{z} = 1/(2A^2) - z$,
by which
\beq
\mu_A = \sqrt{\rho^2 + \bar{z}^2} + \bar{z} \,,
\eeq
and then defining the new constants
\beq
2\alpha_1^2 = \frac{2A^2}{2A^2 z_1 - 1} \,, \quad
2\alpha_2^2 = \frac{2A^2}{2A^2 z_2 - 1} \,.
\eeq
One finally finds
\beq
\label{bonnor}
\psi = -\frac{m_1}{\sqrt{\rho^2 + (\bar{z} - \frac{1}{2\alpha_1^2})^2}} - \frac{m_2}{\sqrt{\rho^2 + (\bar{z} - \frac{1}{2\alpha_2^2})^2}} + \frac{1}{2} \log \bigl( \sqrt{\rho^2 + \bar{z}^2} + \bar{z} \bigr) \,,
\eeq
which is the Bonnor--Swaminarayan potential (cf.~\cite{bonnor1964exact} and~\cite{Griffiths:2009dfa}).
The generalisation to $N$ accelerating particles is easily found as
\beq
\psi = - \sum_{k=1}^N \frac{m_k}{\sqrt{\rho^2 + (\bar{z} - \frac{1}{2\alpha_k^2})^2}} + \frac{1}{2} \log \bigl( \sqrt{\rho^2 + \bar{z}^2} + \bar{z} \bigr) \,.
\eeq
The $\gamma$ function, which completes the Weyl metric~\eqref{weyl}, is found by quadratures.

One can check that the Bonnor--Swaminarayan metric~\eqref{bonnor}, for generic values of the parameters, is affected by conical singularities.
Such singularities can be removed only when $m_2<0$:
in this case the axis is everywhere regular, except at the locations of the point particles.
More explicitly, the regularisation is achieved for~\cite{Griffiths:2009dfa}
\beq
m_1 = -m_2 = \frac{\bigl(\alpha_1^2 - \alpha_2^2)^2}{4\alpha_1^3\alpha_2^3} \,.
\eeq

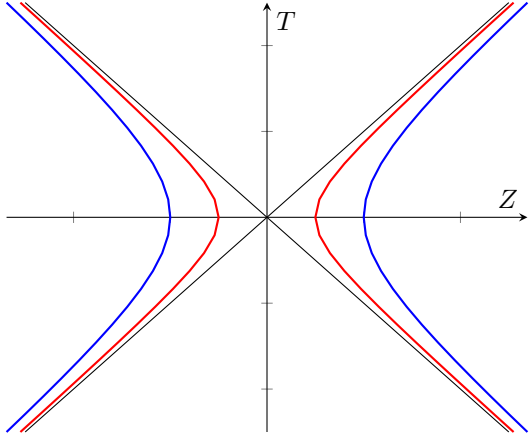
\begin{figure}
\centering
\begin{tikzpicture}

\begin{axis}[
  axis lines=middle,
  clip=false,
  xlabel=$Z$,
  ylabel=$T$,
  xticklabels=\empty,
  yticklabels=\empty,
]

\addplot[black, thin] (x,x);
\addplot[black, thin] (x,-x);

\addplot[blue, thick] (sqrt(x*x+4),x);
\addplot[blue, thick] (-sqrt(x*x+4),x);

\addplot[red, thick] (sqrt(x*x+1),x);
\addplot[red, thick] (-sqrt(x*x+1),x);

\end{axis}

\end{tikzpicture}
\caption{A spacetime diagram of the Bonnor--Swaminarayan in the boost-rotation coordinates of~\cite{Bicak:1999sa}, for the section $\rho=0$.
The hyperbolae represent the worldlines of two pairs of (causally disconnected) accelerating particles.}
\label{fig:hyperbolae}
\end{figure}

\subsection{The limit to the Bi\v c\'ak--Hoenselaers--Schmidt metric}

The inclusion of the external gravitational field is now a simple matter:
this can be done following the lines of section~\ref{sec:ism}, i.e.~by means if the inverse scattering method, or via the same limiting procedure of the multi-black hole metric~\eqref{n-acc} of the previous subsection, now with the field parameters $b_n$ turned on.
In both cases, the resulting Weyl potential is
\beq
\label{bhs}
\psi = - \sum_{k=1}^N \frac{m_k}{\sqrt{\rho^2 + (\bar{z} - \frac{1}{2\alpha_k^2})^2}} + \frac{1}{2} \log \bigl( \sqrt{\rho^2 + \bar{z}^2} + \bar{z} \bigr) + \sum_{n=1}^{\infty} b_n r^n P_n \,.
\eeq
The meaning of the terms in the potential is clear, and the function $\gamma$ can be found again by quadratures.
This potential specialises to the Bi\v c\'ak--Hoenselaers--Schmidt one~\cite{Bicak} for $N=1$ and $b_1\neq0$, $b_n=0$ for $n\geq2$, that reads
\beq
\psi = - \frac{m_1}{\sqrt{\rho^2 + (\bar{z} - \frac{1}{2\alpha_1^2})^2}} + \frac{1}{2} \log \bigl( \sqrt{\rho^2 + \bar{z}^2} + \bar{z} \bigr) + b_1 z \,.
\eeq
The limiting procedure that leads to~\eqref{bhs} does not affect the external field, hence it can be fine tuned again to support the particles attraction against the gravitational collapse and to remove the conical singularities.

It is worth mentioning that Bi\v c\'ak--Hoenselaers--Schmidt also found two accelerating particles described by internal multipole momenta~\cite{BicakMultipole}, in analogy with the Erez--Rosen metric~\cite{erez-rosen}.
The multipole metric, initially obtained through a coalescing limit of the Bonnor--Swaminarayan solution, can be regularised everywhere on the $z$-axis (except at the two particles), thus obtaining a regular accelerating metric without the need of any external field.
Relying on the discussion presented in~\cite{Astorino:2021boj}, one can easily write down the most general potential for $N$ accelerating particles with arbitrary multipole momenta and immersed in an external gravitational field:
\beq
\psi = - \sum_{k=1}^N \frac{m_k}{\sqrt{\rho^2 + (\bar{z} - \frac{1}{2\alpha_k^2})^2}} + \frac{1}{2} \log \bigl( \sqrt{\rho^2 + \bar{z}^2} + \bar{z} \bigr) + \sum_{n=1}^{\infty} \biggl( \frac{a_n}{r^{n+1}} + b_n r^n \biggr) P_n \,.
\eeq
$a_n$ are the internal momenta, that describe the deformations of the point-like sources.
We do not delve into the details of this solution, because it is beyond the scope of this paper.
However, it would be interesting to explicitly write down the function $\gamma$ and to check that the conical singularities can be removed by tuning the parameters $a_n$ and $b_n$.

\section{Summary and Conclusions}

In this article we constructed, thanks to the inverse scattering method, a large family of new solutions which generalise the accelerating Israel--Khan metric~\cite{Dowker:2001dg}, by the introduction of an external gravitational field.
We are able to treat analytically the whole multipolar expansion of the gravitational background introducing a countable number of integration constants, characterising the gravitational multipoles of the external field, which are useful to remove all the conical singularities typical of the accelerating collinear multi-black hole configurations.
As a concrete example, we explicitly analyse some simple configurations such as the accelerating Schwarzschild or the accelerating Bach--Weyl double black hole embedded in an external gravitational field endowed with dipole and quadrupole moments.
We studied the thermodynamic of these solutions and verified the Smarr and the first law.

We find these spacetimes relevant not only because they enrich our scarce theoretical knowledge of multi-black hole solutions or because they are the first accelerating black hole solutions which can be regularised without the need of extra fields as the electromagnetic field\footnote{Actually, in the presence of Maxwell electrodynamics, Ernst showed also how to regularise the charged C-metric, thanks to an external electromagnetic field such as the Melvin universe.
Even though axial magnetic fields in the center of the galaxies can be of some prominence, charged black holes are not considered plausible objects because matter in the universe is most often neutral.}, but also because these metrics allow us to discuss some intriguing physical processes.
In fact regularised C-metrics can describe the pair creation of a couple (or eventually four in case of the double C-metric) of black holes~\cite{pair-bh-grav-field} that accelerate in opposite direction remaining causally disconnected.
This process is propelled at expense of the external field, in our case the multipolar gravitational background.
The significance and the novelty of our picture is given by the fact that the accelerating black hole couple can be uncharged, a feature more in line with phenomenological observation.

As a by-product of our construction we show how to extend the vacuum Pleba\'nski--Demia\'nski class of metrics to the rotating and accelerating multi-Kerr black holes, with or without the presence of the external gravitational field.

We are also able to detect some notable known metrics as limits of our general solution describing accelerating particles with or without the external gravitational background such as the Bonnor–-Swaminarayan and the Bi\v c\'ak--Hoenselaers--Schmidt solutions.

In general we have shown that, as the inverse scattering method predicts, in practice basically any diagonal seed can be used as a background for the solution generating technique. In particular this technique reveals to be useful to embed and overlap a generic number of black hole sources, possibly providing a mechanism to regularise the conical singularities that usually afflicts these metrics. Of course it would be interesting to explore also different backgrounds.

All these results can be extended to gravitational theories where the solution generating techniques hold, from minimally to conformally coupled scalar fields or other scalar tensor theories such as some classes of Brans--Dicke or $f(R)$ gravity\footnote{For details to adapt general relativity solution generating techniques to these other theories see \cite{Astorino:2013xc,Astorino:2014mda}.}.

\section*{Aknowledgments}

{\small This work was supported in part by Conicyt--Beca Chile n$^\textrm{o}$ 74200076, by MIUR-PRIN contract 2017CC72MK003 and also by INFN.}

\bibliographystyle{unsrturl}
\bibliography{RefAcc.bib}

\begin{thebibliography}{10}

\bibitem{Astorino:2021dju}
M.~Astorino and A.~Vigan\`o.
\newblock {Binary black hole system at equilibrium}.
\newblock 2021.
\newblock \href {http://arxiv.org/abs/2104.07686} {\path{arXiv:2104.07686}}.

\bibitem{Astorino:2021boj}
M.~Astorino and A.~Vigan\`o.
\newblock {Charged and rotating multi-black holes in an external gravitational
  field}.
\newblock 2021.
\newblock \href {http://arxiv.org/abs/2105.02894} {\path{arXiv:2105.02894}}.

\bibitem{Israel1964}
W.~Israel and K.A. Khan.
\newblock Collinear particles and bondi dipoles in general relativity.
\newblock {\em Il Nuovo Cimento}, 33(2):331--344, 1964.

\bibitem{NovikovZeldovich}
A.G. Doroshkevich, Y.B. Zel'dovich, and I.D. Novikov.
\newblock {Gravitational Collapse of Non-Symmetric and Rotating Bodies}.
\newblock {\em Zhurnal Eksperimentalnoi i Teoreticheskoi Fiziki}, 49:170,
  December 1965.

\bibitem{Chandrasekhar:1985kt}
S.~Chandrasekhar.
\newblock {\em {The mathematical theory of black holes}}.
\newblock Clarendon Press, Oxford, 1985.

\bibitem{Geroch:1982bv}
R.P. Geroch and J.B. Hartle.
\newblock {Distorted black holes}.
\newblock {\em J. Math. Phys.}, 23:680, 1982.
\newblock \href {https://doi.org/10.1063/1.525384}
  {\path{doi:10.1063/1.525384}}.

\bibitem{breton-manko}
N.~Breton, A.A. Garcia, V.S. Manko, and T.E. Denisova.
\newblock {Arbitrarily deformed Kerr--Newman black hole in an external
  gravitational field}.
\newblock {\em Phys. Rev. D}, 57:3382--3388, 1998.
\newblock \href {https://doi.org/10.1103/PhysRevD.57.3382}
  {\path{doi:10.1103/PhysRevD.57.3382}}.

\bibitem{Abdolrahimi:2015gea}
S.~Abdolrahimi, J.~Kunz, P.~Nedkova, and C.~Tzounis.
\newblock {Properties of the distorted Kerr black hole}.
\newblock {\em JCAP}, 12:009, 2015.
\newblock \href {http://arxiv.org/abs/1509.01665} {\path{arXiv:1509.01665}},
  \href {https://doi.org/10.1088/1475-7516/2015/12/009}
  {\path{doi:10.1088/1475-7516/2015/12/009}}.

\bibitem{Fairhurst:2000xh}
S.~Fairhurst and B.~Krishnan.
\newblock {Distorted black holes with charge}.
\newblock {\em Int. J. Mod. Phys. D}, 10:691--710, 2001.
\newblock \href {http://arxiv.org/abs/gr-qc/0010088}
  {\path{arXiv:gr-qc/0010088}}, \href
  {https://doi.org/10.1142/S0218271801001086}
  {\path{doi:10.1142/S0218271801001086}}.

\bibitem{ernst}
F.J. Ernst.
\newblock Generalized c‐metric.
\newblock {\em Journal of Mathematical Physics}, 19(9):1986--1987, 1978.
\newblock \href {https://doi.org/https://doi.org/10.1063/1.523896}
  {\path{doi:https://doi.org/10.1063/1.523896}}.

\bibitem{Gibbons:1974zd}
G.W. Gibbons.
\newblock {The Motion of black holes}.
\newblock {\em Commun. Math. Phys.}, 35:13--23, 1974.
\newblock \href {https://doi.org/10.1007/BF01646451}
  {\path{doi:10.1007/BF01646451}}.

\bibitem{Belinsky:1971nt}
V.A. Belinsky and V.E. Zakharov.
\newblock {Integration of the Einstein Equations by the Inverse Scattering
  Problem Technique and the Calculation of the Exact Soliton Solutions}.
\newblock {\em Sov. Phys. JETP}, 48:985--994, 1978.

\bibitem{Belinsky:1979mh}
V.A. Belinsky and V.E. Sakharov.
\newblock {Stationary Gravitational Solitons with Axial Symmetry}.
\newblock {\em Sov. Phys. JETP}, 50:1--9, 1979.

\bibitem{Belinski:2001ph}
V.~Belinski and E.~Verdaguer.
\newblock {\em {Gravitational solitons}}.
\newblock Cambridge Monographs on Mathematical Physics. Cambridge University
  Press, 2005.
\newblock \href {https://doi.org/10.1017/CBO9780511535253}
  {\path{doi:10.1017/CBO9780511535253}}.

\bibitem{Weyl1917}
H.~Weyl.
\newblock Zur gravitationstheorie.
\newblock {\em Annalen der Physik}, 359(18):117--145, 1917.
\newblock \href {https://doi.org/https://doi.org/10.1002/andp.19173591804}
  {\path{doi:https://doi.org/10.1002/andp.19173591804}}.

\bibitem{Dowker:2001dg}
H.F. Dowker and S.N. Thambyahpillai.
\newblock {Many accelerating black holes}.
\newblock {\em Class. Quant. Grav.}, 20:127--136, 2003.
\newblock \href {http://arxiv.org/abs/gr-qc/0105044}
  {\path{arXiv:gr-qc/0105044}}, \href
  {https://doi.org/10.1088/0264-9381/20/1/310}
  {\path{doi:10.1088/0264-9381/20/1/310}}.

\bibitem{deCastro:2011zz}
G.M. de~Castro and P.S. Letelier.
\newblock {Black holes surrounded by thin rings and the stability of circular
  orbits}.
\newblock {\em Class. Quant. Grav.}, 28:225020, 2011.
\newblock \href {https://doi.org/10.1088/0264-9381/28/22/225020}
  {\path{doi:10.1088/0264-9381/28/22/225020}}.

\bibitem{LetelierBrasil}
P.S. Letelier.
\newblock Multipole stationary soliton solutions to the einstein equations.
\newblock {\em Revista Brasileira de Fisica}, 14:371--376, September 1984.

\bibitem{Hong:2004dm}
K.~Hong and E.~Teo.
\newblock {A New form of the rotating C-metric}.
\newblock {\em Class. Quant. Grav.}, 22:109--118, 2005.
\newblock \href {http://arxiv.org/abs/gr-qc/0410002}
  {\path{arXiv:gr-qc/0410002}}, \href
  {https://doi.org/10.1088/0264-9381/22/1/007}
  {\path{doi:10.1088/0264-9381/22/1/007}}.

\bibitem{LetelierSolitons}
P.S. Letelier.
\newblock Static and stationary multiple soliton solutions to the einstein
  equations.
\newblock {\em Journal of Mathematical Physics}, 26(3):467--476, 1985.
\newblock \href {https://doi.org/10.1063/1.526633}
  {\path{doi:10.1063/1.526633}}.

\bibitem{Griffiths:2006tk}
J.B. Griffiths, P.~Krtous, and J.~Podolsky.
\newblock {Interpreting the C-metric}.
\newblock {\em Class. Quant. Grav.}, 23:6745--6766, 2006.
\newblock \href {http://arxiv.org/abs/gr-qc/0609056}
  {\path{arXiv:gr-qc/0609056}}, \href
  {https://doi.org/10.1088/0264-9381/23/23/008}
  {\path{doi:10.1088/0264-9381/23/23/008}}.

\bibitem{Hong:2003gx}
K.~Hong and E.~Teo.
\newblock {A New form of the C metric}.
\newblock {\em Class. Quant. Grav.}, 20:3269--3277, 2003.
\newblock \href {http://arxiv.org/abs/gr-qc/0305089}
  {\path{arXiv:gr-qc/0305089}}, \href
  {https://doi.org/10.1088/0264-9381/20/14/321}
  {\path{doi:10.1088/0264-9381/20/14/321}}.

\bibitem{Komar}
A.~Komar.
\newblock Covariant conservation laws in general relativity.
\newblock {\em Phys. Rev.}, 113:934--936, Feb 1959.
\newblock \href {https://doi.org/10.1103/PhysRev.113.934}
  {\path{doi:10.1103/PhysRev.113.934}}.

\bibitem{Tomimatsu:1984pw}
A.~Tomimatsu.
\newblock {Equilibrium of Two Rotating Charged Black Holes and the Dirac
  String}.
\newblock {\em Prog. Theor. Phys.}, 72:73, 1984.
\newblock \href {https://doi.org/10.1143/PTP.72.73}
  {\path{doi:10.1143/PTP.72.73}}.

\bibitem{Gregory:2020mmi}
R.~Gregory, Z.L. Lim, and A.~Scoins.
\newblock {Thermodynamics of Many Black Holes}.
\newblock {\em Front. in Phys.}, 9:187, 2021.
\newblock \href {http://arxiv.org/abs/2012.15561} {\path{arXiv:2012.15561}},
  \href {https://doi.org/10.3389/fphy.2021.666041}
  {\path{doi:10.3389/fphy.2021.666041}}.

\bibitem{Harmark:2004rm}
T.~Harmark.
\newblock {Stationary and axisymmetric solutions of higher-dimensional general
  relativity}.
\newblock {\em Phys. Rev. D}, 70:124002, 2004.
\newblock \href {http://arxiv.org/abs/hep-th/0408141}
  {\path{arXiv:hep-th/0408141}}, \href
  {https://doi.org/10.1103/PhysRevD.70.124002}
  {\path{doi:10.1103/PhysRevD.70.124002}}.

\bibitem{bonnor1988exact}
W.B. Bonnor.
\newblock An exact solution of einstein's equations for two particles falling
  freely in an external gravitational field.
\newblock {\em General relativity and gravitation}, 20(6):607--622, 1988.
\newblock \href {https://doi.org/https://doi.org/10.1007/BF00758917}
  {\path{doi:https://doi.org/10.1007/BF00758917}}.

\bibitem{Astorino:2013xxa}
M.~Astorino.
\newblock {Pair Creation of Rotating Black Holes}.
\newblock {\em Phys. Rev. D}, 89(4):044022, 2014.
\newblock \href {http://arxiv.org/abs/1312.1723} {\path{arXiv:1312.1723}},
  \href {https://doi.org/10.1103/PhysRevD.89.044022}
  {\path{doi:10.1103/PhysRevD.89.044022}}.

\bibitem{Gibbons:1986cq}
G.W. Gibbons.
\newblock {Quantized flux tubes in Einstein-Maxwell theory and noncompact
  internal spaces}.
\newblock In {\em {22nd Winter School of Theoretical Physics: Fields and
  Geometry}}, 5 1986.

\bibitem{strominger}
D.~Garfinkle, S.B. Giddings, and A.~Strominger.
\newblock {Entropy in black hole pair production}.
\newblock {\em Phys. Rev. D}, 49:958--965, 1994.
\newblock \href {http://arxiv.org/abs/gr-qc/9306023}
  {\path{arXiv:gr-qc/9306023}}, \href {https://doi.org/10.1103/PhysRevD.49.958}
  {\path{doi:10.1103/PhysRevD.49.958}}.

\bibitem{hawking}
S.W. Hawking, G.T. Horowitz, and S.F. Ross.
\newblock {Entropy, Area, and black hole pairs}.
\newblock {\em Phys. Rev. D}, 51:4302--4314, 1995.
\newblock \href {http://arxiv.org/abs/gr-qc/9409013}
  {\path{arXiv:gr-qc/9409013}}, \href
  {https://doi.org/10.1103/PhysRevD.51.4302}
  {\path{doi:10.1103/PhysRevD.51.4302}}.

\bibitem{bonnor}
W.B. Bonnor.
\newblock Static magnetic fields in general relativity.
\newblock {\em Proceedings of the Physical Society. Section A}, 67(3):225,
  1954.
\newblock \href {https://doi.org/https://doi.org/10.1088/0370-1298/67/3/305}
  {\path{doi:https://doi.org/10.1088/0370-1298/67/3/305}}.

\bibitem{melvin}
M.A. Melvin.
\newblock {Pure magnetic and electric geons}.
\newblock {\em Phys. Lett.}, 8:65--70, 1964.
\newblock \href {https://doi.org/10.1016/0031-9163(64)90801-7}
  {\path{doi:10.1016/0031-9163(64)90801-7}}.

\bibitem{pair-bh-grav-field}
M.~Astorino and A.~Vigan\`o.
\newblock Pair creation of black holes in gravitational background.
\newblock {\em In preparation}.

\bibitem{Christodoulou:1972kt}
D.~Christodoulou and R.~Ruffini.
\newblock {Reversible transformations of a charged black hole}.
\newblock {\em Phys. Rev. D}, 4:3552--3555, 1971.
\newblock \href {https://doi.org/10.1103/PhysRevD.4.3552}
  {\path{doi:10.1103/PhysRevD.4.3552}}.

\bibitem{Curzon}
H.E.J. Curzon.
\newblock Cylindrical solutions of einstein's gravitation equations.
\newblock {\em Proceedings of the London Mathematical Society}, 2(1):477--480,
  1925.
\newblock \href {https://doi.org/https://doi.org/10.1112/plms/s2-23.1.477}
  {\path{doi:https://doi.org/10.1112/plms/s2-23.1.477}}.

\bibitem{Chazy}
J.~Chazy.
\newblock Sur le champ de gravitation de deux masses fixes dans la th{\'e}orie
  de la relativit{\'e}.
\newblock {\em Bulletin de la Societe mathematique de France}, 52:17--38, 1924.
\newblock \href {https://doi.org/https://doi.org/10.24033/bsmf.1044}
  {\path{doi:https://doi.org/10.24033/bsmf.1044}}.

\bibitem{Einstein:1936fp}
A.~Einstein and N.~Rosen.
\newblock {Two-Body Problem in General Relativity Theory}.
\newblock {\em Phys. Rev.}, 49:404--405, 1936.
\newblock \href {https://doi.org/10.1103/PhysRev.49.404.2}
  {\path{doi:10.1103/PhysRev.49.404.2}}.

\bibitem{bonnor1964exact}
W.B. Bonnor and N.S. Swaminarayan.
\newblock An exact solution for uniformly accelerated particles in general
  relativity.
\newblock {\em Zeitschrift f{\"u}r Physik}, 177(3):240--256, 1964.
\newblock \href {https://doi.org/https://doi.org/10.1007/BF01375497}
  {\path{doi:https://doi.org/10.1007/BF01375497}}.

\bibitem{Bicak}
J.~Bicak, C.~Hoenselaers, and B.G. Schmidt.
\newblock The solutions of the einstein equations for uniformly accelerated
  particles without nodal singularities. i. freely falling particles in
  external fields.
\newblock {\em Proceedings of the Royal Society of London. A. Mathematical and
  Physical Sciences}, 390(1799):397--409, 1983.
\newblock \href {https://doi.org/10.1098/rspa.1983.0138}
  {\path{doi:10.1098/rspa.1983.0138}}.

\bibitem{Griffiths:2009dfa}
J.B. Griffiths and J.~Podolsky.
\newblock {\em {Exact Space-Times in Einstein's General Relativity}}.
\newblock Cambridge Monographs on Mathematical Physics. Cambridge University
  Press, Cambridge, 2009.
\newblock \href {https://doi.org/10.1017/CBO9780511635397}
  {\path{doi:10.1017/CBO9780511635397}}.

\bibitem{Bicak:1999sa}
J.~Bicak and V.~Pravda.
\newblock {Spinning C metric: Radiative space-time with accelerating, rotating
  black holes}.
\newblock {\em Phys. Rev. D}, 60:044004, 1999.
\newblock \href {http://arxiv.org/abs/gr-qc/9902075}
  {\path{arXiv:gr-qc/9902075}}, \href
  {https://doi.org/10.1103/PhysRevD.60.044004}
  {\path{doi:10.1103/PhysRevD.60.044004}}.

\bibitem{BicakMultipole}
J.~Bicak, C.~Hoenselaers, and B.G. Schmidt.
\newblock The solutions of the einstein equations for uniformly accelerated
  particles without nodal singularities. ii. self-accelerating particles.
\newblock {\em Proceedings of the Royal Society of London. A. Mathematical and
  Physical Sciences}, 390(1799):411--419, 1983.
\newblock \href {https://doi.org/10.1098/rspa.1983.0139}
  {\path{doi:10.1098/rspa.1983.0139}}.

\bibitem{erez-rosen}
G.~Erez and N.~Rosen.
\newblock The gravitational field of a particle possessing a multipole moment.
\newblock {\em Bull. Research Council Israel}, Sect. F.8, 9 1959.

\bibitem{Astorino:2013xc}
M.~Astorino.
\newblock {Embedding hairy black holes in a magnetic universe}.
\newblock {\em Phys. Rev. D}, 87(8):084029, 2013.
\newblock \href {http://arxiv.org/abs/1301.6794} {\path{arXiv:1301.6794}},
  \href {https://doi.org/10.1103/PhysRevD.87.084029}
  {\path{doi:10.1103/PhysRevD.87.084029}}.

\bibitem{Astorino:2014mda}
M.~Astorino.
\newblock {Stationary axisymmetric spacetimes with a conformally coupled scalar
  field}.
\newblock {\em Phys. Rev. D}, 91:064066, 2015.
\newblock \href {http://arxiv.org/abs/1412.3539} {\path{arXiv:1412.3539}},
  \href {https://doi.org/10.1103/PhysRevD.91.064066}
  {\path{doi:10.1103/PhysRevD.91.064066}}.

\end{thebibliography}

\end{document}